# Robust Model Predictive Control for Autonomous Vehicle/Self-Driving Cars

Che Kun Law, Darshit Dalal, Stephen Shearrow

A robust Model Predictive Control (MPC) approach for controlling front steering of an autonomous vehicle is presented in this project. we present various approaches to increase the robustness of model predictive control by using weight tuning, a successive on-line linearization of a nonlinear vehicle model similar to the approach in [1] and successive on-line linearization with a velocity model.

## I. Nomenclature

| | | |
|---|---|---|
| $a$ | = | scaling factor used in weight tuning |
| $\beta$ | = | angle of the current velocity of the center of mass w.r.t. the longitudinal axis of the car |
| $l_f$ | = | length from front wheels to center of mass |
| $l_r$ | = | length from rear wheels to center of mass |
| $T_s$ | = | sampling time |
| V | = | speed of the vehicle |
| x | = | position along x-axis |
| y | = | position along y-axis |
| $\psi$ | = | heading angle of the vehicle |
| $\delta_f$ | = | steering angle of front wheel |
| ' | = | rate change of variable |

## II. Introduction

Work in autonomous vehicles has grown exponentially in the past few years due to advances in computing power and sensing technologies. The main components of a modern autonomous vehicle/self-driving car are localization, perception, and control. This report will discuss the control of the vehicle's steering using Model Predictive Control (MPC). In MPC [2], at each sampling time step, starting at the current state, an open-loop optimal control problem is solved over a finite prediction horizon. The optimal control input is applied to the system to propagate the system dynamics. At the next time step, a new optimal control problem is solved over a shifted horizon based on the current state of the system. The optimal solution requires a dynamic model of the system, enables input and output constraints to be explicitly included and minimizes a cost function.

Model Predictive Control is susceptible to modeling uncertainties, as well as change in model dynamics due to non-linearities; so we seek to make our controller more robust by introducing elements such as weight tuning, successive linearization, and successive linearization with a velocity model. This will enable our controller to be more robust and better meet the safety standards of the automotive industry.

In our problem we have assumed that a path-planning algorithm has already generated an optimal trajectory for our autonomous vehicle to follow. The goal of our vehicle is then to track this trajectory with the least error while satisfying our constraints such as the rate of steering angle. We utilize the nonlinear kinematic bicycle model as described in [3] for the dynamics of the vehicle. The kinematic model is more computationally efficient than a dynamic vehicle model with a tire model. Also, it avoids the tire model being singular at low vehicle speeds, as tire models have a tire slip angle estimation term which has vehicle velocity in the denominator. It has also been shown to have comparable performance to a higher fidelity vehicle model.

## III. Dynamics

$$\begin{bmatrix} \dot{x} \\ \dot{y} \\ \dot{\psi} \\ \dot{v} \\ \dot{\beta} \end{bmatrix} = \begin{bmatrix} v\cos(\psi + \beta) \\ v\sin(\psi + \beta) \\ \dfrac{v}{l_r}\sin(\beta) \\ a \\ \tan^{-1}\left(\dfrac{l_r}{l_f + l_r}\tan(\delta_f)\right) \end{bmatrix} \quad (1)$$

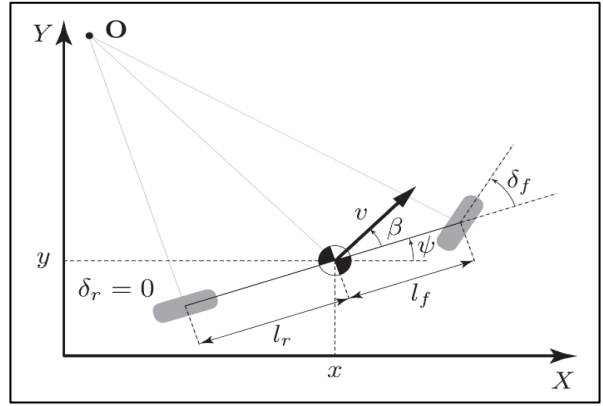

**Figure 1: Kinematic Bicycle Model**

The dynamics of our system can be seen in figure 1, in our model we assume v to be constant and our control input is the rate of change of β. This can be mapped onto a corresponding steering angle given the relationship between β and $\delta_f$ as shown in figure 1. The constraint we apply on our system is a constraint on the rate of change of steering angle, we formulate this by constraining the rate of change of beta is which directly correlated to steering angle.

## IV. Weight Tuning

### A. Formulation

In this section, weight tuning is used as a first cut controller design for MPC. This tuning method provides insight on appropriate weights to be used in later controller iterations given input constraints. The weights directly impact the cost function and in turn impact the performance of the controller. In this method, the bicycle model is linearized at the initial point and is used as the internal model for the MPC controller.

Since the application of this controller is a self-driving car, a few assumptions and constraints are put on the system. To ensure driver comfort, a rate limit of 0.5 rad/sec is put on the steering angle. This is approximately 30 degrees/sec. It is assumed that the vehicle will be driving at a constant speed of 10 m/s on a flat surface.

To set up this problem, the bicycle model is discretized then linearized as shown below. The equations are linearized about the initial point using the small angle approximation.

Discretize:

$$x(k + 1) = x(k) + V\cos(\varphi(k) + \beta(k))T_s \quad (2)$$

$$(k + 1) = y(k) + V\sin(\varphi(k) + \beta(k))T_s \quad (3)$$

$$\varphi(k+1) = \varphi(k) + \frac{V}{lr}\sin(\beta(k))T_s \tag{4}$$

$$\beta = \tan^{-1}\left(\frac{lr}{lf+lr}\tan(\delta_f)\right) \tag{5}$$

Linearize:

$$x(k+1) = x(k) + V * T_s \tag{6}$$

$$y(k+1) = y(k) + V * (\varphi(k) + \beta(k)) * T_s \tag{7}$$

$$\varphi(k+1) = \varphi(k) + \frac{V}{lr} * \beta(k) * T_s \tag{8}$$

A state-space model is formed using the linearized and discretized equations.

$$x(k+1) = \begin{bmatrix} 1 & 0 & 0 \\ 0 & 1 & TsV \\ 1 & 0 & 0 \end{bmatrix}\begin{bmatrix} x(k) \\ y(k) \\ \varphi(k) \end{bmatrix} + \begin{bmatrix} 0 \\ TsV \\ TsV/lr \end{bmatrix}\beta + \begin{bmatrix} VTs \\ 0 \\ 0 \end{bmatrix} \tag{9}$$

Weights are chosen for the manipulated variable, manipulated variable rate, and output variable. These weights are 0, 0.1, and 10 respectfully. These weights directly impact the cost function and are chosen to maximize output tracking performance. Next, the following parameters are used as defaults when first synthesizing a controller.

| Variable [units] | Value |
|---|---|
| Sampling Time [Seconds] | 0.2 |
| Prediction Horizon [Time Steps] | 10 |
| Control Horizon [Time Steps] | 5 |
| Manipulated Variable Weight [$W_{i,j}^u$] | 0 |
| Manipulated Variable Rate Weight [$W_{i,j}^{\Delta u}$] | 0.1 |
| Output Variable Weight [$W_{i,j}^y$] | 10 |

**Table 1: Default Parameters**

Below are the cost functions associated with each weight.

Output Reference Tracking

$$J_y(z_k) = \sum_{j=1}^{n_y}\sum_{i=1}^{p}\left\{W_{i,j}^y * [r_j(k+i|k) - y_j(k+i|k)]\right\}^2 \tag{10}$$

Manipulated Variable Tracking

$$J_u(z_k) = \sum_{j=1}^{n_u}\sum_{i=0}^{p-1}\left\{W_{i,j}^u * [u_j(k+i|k) - u_{j,target}(k+i|k)]\right\}^2 \tag{11}$$

Manipulated Variable Move Suppression

$$J_{\Delta u}(z_k) = \sum_{j=1}^{n_u}\sum_{i=0}^{p-1}\left\{W_{i,j}^{\Delta u} * [u_j(k+i|k) - u_j(k+i-1|k)]\right\}^2 \tag{12}$$

Using the state-space model and above parameters, MATLAB [4] is used to synthesize and simulate various controller designs. To simplify tuning, the α variable is used to scale all the weights at the same time. This variable impacts the aggressiveness and conservativeness of the controller. The relationship between α and the weights are shown below.

$$\text{Scaled } W_{i,j}^{y} = W_{i,j}^{y} * \alpha \tag{13}$$

$$\text{Scaled } W_{i,j}^{u} = \frac{W_{i,j}^{u}}{\alpha} \tag{14}$$

$$\text{Scaled } W_{i,j}^{\Delta u} = W_{i,j}^{\Delta u} * \alpha \tag{15}$$

### B. Simulations and Results

Several controllers were synthesized with varying $\alpha$ and a step response simulation was run. Below are the results.

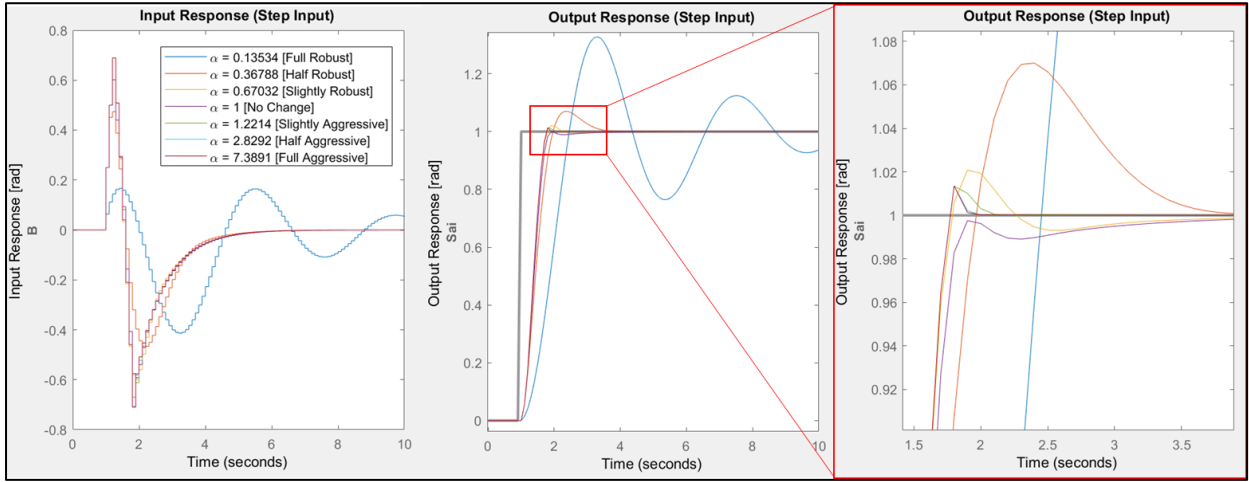

**Figure 2: Step Response Simulation**

*Note: These responses are simulated with no disturbance or noise using the default parameters. α is the only variable that changes.*

As α is increased, the controller is more aggressive in achieving the desired state. Conversely, as α is decreased the controller is more relaxed in achieving the desired state as shown in the plots above. Next, gaussian output disturbance is added to the system with an amplitude of 0.05. This amplitude is chosen to simulate gusts of wind while the car is driving.

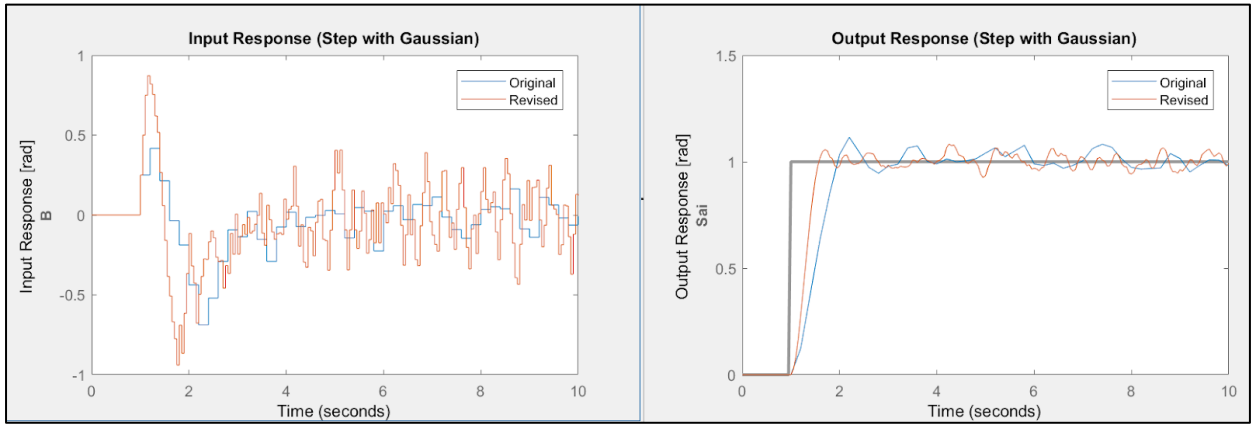

**Figure 3: Step Response with Gaussian Disturbance Simulation**

Using the half aggressive controller shown in the previous simulation with $\alpha = 2.8$ as the original, we see that with gaussian output disturbance, the controller tracks a step response well. However, adjusting the prediction horizon and sampling time will improve the response. The revised controller has a sampling time of 0.05 and a prediction horizon of twenty time steps. The next simulation uses a sine wave as the reference trajectory while still having gaussian output disturbance.

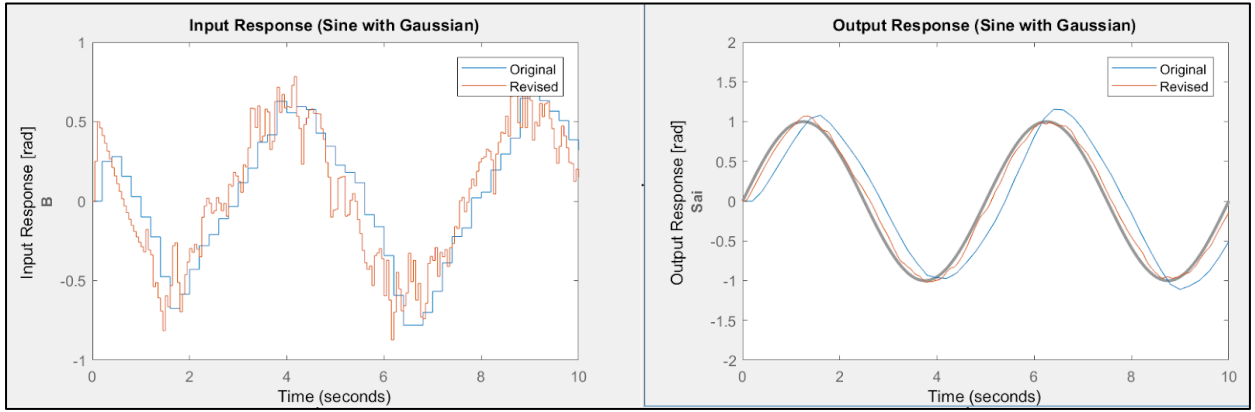

**Figure 4: Sine Wave Response with Gaussian Disturbance Simulation**

From this simulation, the revised controller tracks the sine wave significantly better than the original controller. The revised MPC controller will be used as a baseline to compare other controller design methods moving forward.

## V. Adaptive Optimal Control using Successive Linearization in a Position Model

We have decided to tackle the nonlinear dynamics of our vehicle model by using successive linearization to linearize the nonlinear vehicle model at each time step and update the model to a new operating point. This enables a linear system to be obtained, and using a quadratic cost function, it can be formulated as a quadratic programming problem which can be solved computationally efficiently.

The system was first discretized using a forward difference method as described in [2]

Given CT model
$$\dot{x}(t) = g^c(x^c(t), u^c(t)) \\ y(t) = h^c(x(t), u^c(t)) \tag{16}$$

Approximate
$$\frac{d}{dt}x^c(t) = \frac{x^c(t + T_s) - x^c(t)}{T_s} \tag{17}$$

$$\xrightarrow{yields} \frac{d}{dt}x^c(t) = \frac{x^c(t + T_s) - x^c(t)}{T_s} \tag{18}$$

Then the discretized model is
$$X(k + 1) = X(k) + T_s g^c(X(k), U(k)) = g(X(k), U(k)) \tag{19}$$

$$Y(k) = h^c(X(k), U(k)) = h(X(k), U(k)) \tag{20}$$

If $T_s$ is small and CT and DT have the "same" initial conditions and inputs, then outputs of CT and DT systems "will be close". After discretizing the system, we apply Taylor Series expansion to linearize the nonlinear system. We take the Taylor Series expansion about the non-linear, X(k) is already known from the dynamics previously.

$$X(k+1) = X(k) + T_s g^c(X(k), u(k)) \tag{21}$$

$$X(k+1) = \begin{bmatrix} x(k) \\ y(k) \\ \psi(k) \end{bmatrix} + G \tag{22}$$

Where G is defined below

$$T_s \begin{bmatrix} v\cos(\psi_o(k) + \beta_o(k)) - v\sin(\psi_o(k) + \beta_o(k))[\psi(k) - \psi_o] - v\sin(\psi_o(k) + \beta_o(k))[\beta(k) - \beta_o] \\ v\sin(\psi_o(k) + \beta_o(k)) + v\cos(\psi_o(k) + \beta_o(k))[\psi(k) - \psi_o] - v\sin(\psi_o(k) + \beta_o(k))[\beta(k) - \beta_o] \\ \frac{v}{l_r}\sin(\beta_o) + \frac{v}{l_r}\cos(\beta_o)[\beta(k) - \beta_o] \end{bmatrix} \tag{23}$$

Assume $\Delta\psi(k) = 0$ for $T_s$

$$G = T_s \begin{bmatrix} v\cos(\psi_o(k) + \beta_o(k)) \\ v\sin(\psi_o(k) + \beta_o(k)) \\ \frac{v}{l_r}\sin(\beta_o) \end{bmatrix} + T_s \begin{bmatrix} -v\sin(\psi_o(k) + \beta_o(k))[\beta(k) - \beta_o] \\ -v\sin(\psi_o(k) + \beta_o(k))[\beta(k) - \beta_o] \\ \frac{v}{l_r}\cos(\beta_o)[\beta(k) - \beta_o] \end{bmatrix} \tag{24}$$

$$X(k+1) = \begin{bmatrix} x(k) \\ y(k) \\ \psi(k) \end{bmatrix} + T_s \begin{bmatrix} v\cos(\psi_o(k) + \beta_o(k)) \\ v\sin(\psi_o(k) + \beta_o(k)) \\ \frac{v}{l_r}\sin(\beta_o) \end{bmatrix}$$
$$+ T_s \begin{bmatrix} -v\sin(\psi_o(k) + \beta_o(k))[\beta(k) - \beta_o] \\ -v\sin(\psi_o(k) + \beta_o(k))[\beta(k) - \beta_o] \\ \frac{v}{l_r}\cos(\beta_o)[\beta(k) - \beta_o] \end{bmatrix} \tag{25}$$

$$\begin{bmatrix} x(k+1) \\ y(k+1) \\ \psi(k+1) \end{bmatrix} = \begin{bmatrix} 1 & 0 & 0 \\ 0 & 1 & 0 \\ 0 & 0 & 1 \end{bmatrix} \begin{bmatrix} x(k) \\ y(k) \\ \psi(k) \end{bmatrix}$$
$$+ \begin{bmatrix} -v\sin(\psi_o + \beta_o) & 0 & 0 \\ 0 & v\cos(\psi_o + \beta_o) & 0 \\ 0 & 0 & v\cos(\beta_o) \end{bmatrix} \Delta\beta T_s \tag{26}$$
$$+ \begin{bmatrix} v\cos(\psi_o + \beta_o) \\ v\sin(\psi_o + \beta_o) \\ v\cos(\beta_o) \end{bmatrix} T_s$$

$$\begin{bmatrix} x(k+1) \\ y(k+1) \\ \psi(k+1) \end{bmatrix} = A \begin{bmatrix} x(k) \\ y(k) \\ \psi(k) \end{bmatrix} + Bu(k)T_s + KT_s \tag{27}$$

*Where A, B, K are matrices, K is a constant matrix added to the typical state space model of the form x(k+1) = Ax(k) + Bu(k)*

The MPC problem can then be formulated with the following cost function

$$\min_u J = \sum_{k}^{N} \left(X(k) - X_{ref}(k)\right)^T \cdot Q \cdot \left(X(k) - X_{ref}(k)\right) + u(k) \cdot R \cdot u(k) \tag{28}$$

$$s.t. \quad u_{min} \leq u \leq u_{max}$$

```
Input: t, x_ref, x_o, N, T_s, E, W
Output: yMPC, uMPC
        Linearize model at initial condition x(0)
        for k = 1 to kmax do
Find u*(k) with mpcqpsolver
Update nonlinear dynamics with 1st input of u*(k): x(k+1) = g(x(k), u*(k))
Update linearized model with new operating point/states
end for
return yMPC, uMPC
```

**Figure 5: Algorithm for Successive Linearization**

## VI. Adaptive Optimal Control using Successive Linearization in a Velocity Model

Velocity Model for successive linearization tracks the velocity error or error in 'change of states in a sampling period' rather than the error in the position itself. The velocity model offers convenience in its expression in state space. It does not need a constant matrix for successive linearization, unlike the Position Model. This further improves the computation time.

### A. Linearization and Discretization

The dynamics for the velocity model are discretized as follows. The states ($\Delta x, \Delta y, \Delta \psi$) are expressed as differences and the control input ($\Delta \beta$) is expressed as a change in the tire angle.

$$\Delta x(k+1) = \Delta x(k) - V.\sin(\psi_o(k) + \beta_o(k)).(\Delta \psi(k) + \Delta \beta(k)).T_s \tag{29}$$

$$\Delta y(k+1) = \Delta y(k) + V.\cos(\psi_o(k) + \beta_o(k)).(\Delta \psi(k) + \Delta \beta(k)).T_s \tag{30}$$

$$\Delta \psi(k+1) = \Delta \psi(k) + (\frac{V}{l_r}).\cos(\beta_o(k)).(\Delta \beta(k)).T_s \tag{31}$$

The equations shown here are already linear in nature as the integrated states (x,y,$\psi$) are assumed to remain constant for the duration of a sampling period. The state space representation is as shown here.

$$\begin{bmatrix} \Delta x(k+1) \\ \Delta y(k+1) \\ \Delta \psi(k+1) \end{bmatrix} = \begin{bmatrix} 1 & 0 & -V.\sin(\psi_o(k) + \beta_o(k).T_s \\ 0 & 1 & V.\cos(\psi_o(k) + \beta_o(k).T_s \\ 0 & 0 & 1 \end{bmatrix} . \begin{bmatrix} \Delta x(k) \\ \Delta y(k) \\ \Delta \psi(k) \end{bmatrix} \\ + \begin{bmatrix} -V.\sin(\psi_o(k) + \beta_o(k).T_s \\ V.\cos(\psi_o(k) + \beta_o(k).T_s \\ (\frac{V}{l_r}).\cos(\beta_o(k).T_s \end{bmatrix} . \Delta \beta(k) \tag{32}$$

### B. Problem Formulation

The MPC problem can be formulated as:

$$\min_u J = \sum_k^N \left(\Delta X(k) - \Delta X_{ref}(k)\right)^T . Q . \left(\Delta X(k) - \Delta X_{ref}(k)\right) + u(k).R.u(k) \tag{33}$$

$$s.t. \quad u_{min} \leq u \leq u_{max}$$

Since the control input is a difference input hence the constraints on the input automatically become rate constraints.

**C. Problem Solution**

The problem can be solved using MATLAB's Model Predictive Control (MPC) Quadratic Programming (QP) solver. The general algorithm for solving the problem can be outlined as follows.

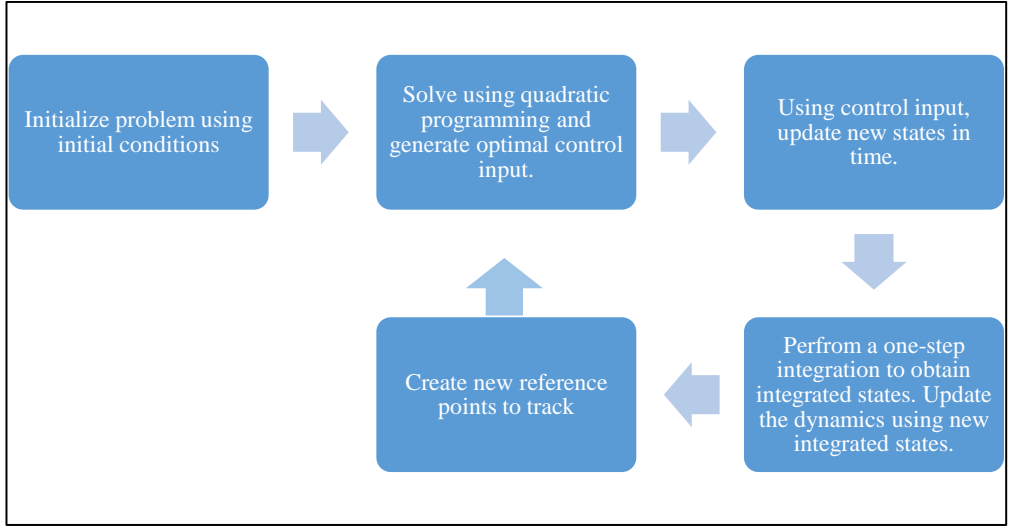

**Figure 6: General Algorithm for Solving the Velocity Model**

**D. Creating New Reference Points**

Generally, input reference trajectories are provided in the form of x and y positions that a controller must track. A controller based on the velocity model must create its own reference points in terms of $\Delta x$ and $\Delta y$. The controller must estimate, the reference point it would be closest to after one sampling period. Based on this estimated future reference point the controller then generates the new $\Delta x_{ref}$ and $\Delta y_{ref}$.

$$x(k+1)_{estimated-reference} = x(k)_{reference} + V.\cos(\psi(k) + \beta(k)).T_s \quad (34)$$

$$y(k+1)_{estimated-reference} = y(k)_{reference} + V.\sin(\psi(k) + \beta(k)).T_s \quad (35)$$

$$\Delta x(k+1)_{reference} = x(k+1)_{estimated-reference} - x(k)_{reference} \quad (36)$$

$$\Delta y(k+1)_{reference} = y(k+1)_{estimated-reference} - y(k)_{reference} \quad (37)$$

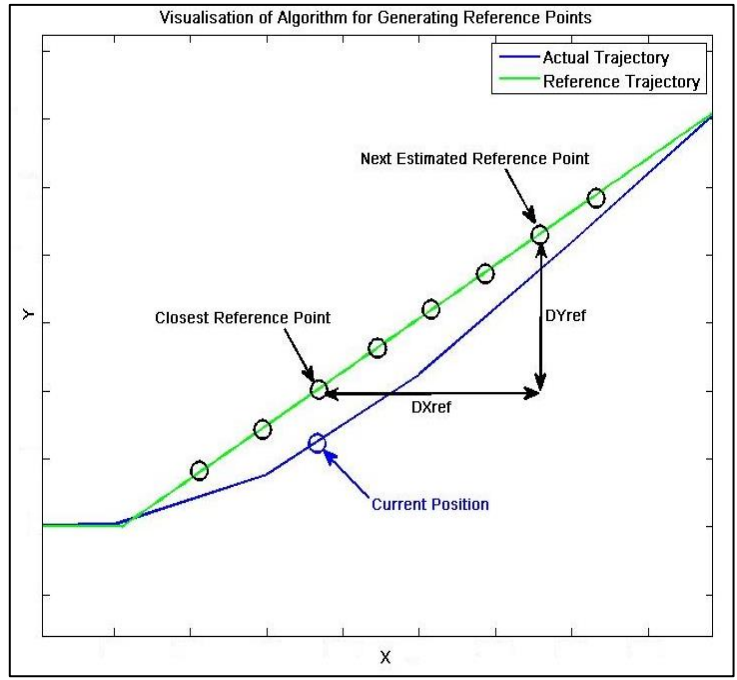

Figure 7: Visualization of Algorithm for Generating Reference Points

## VII. Results

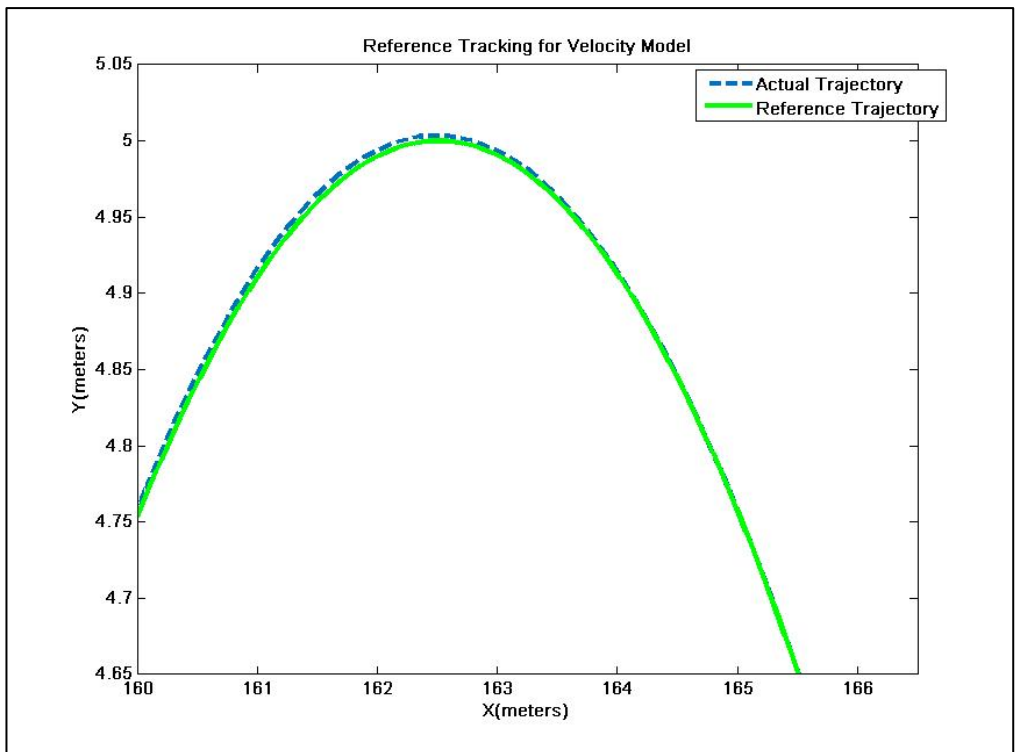

Figure 8a: Reference Tracking for Velocity Model

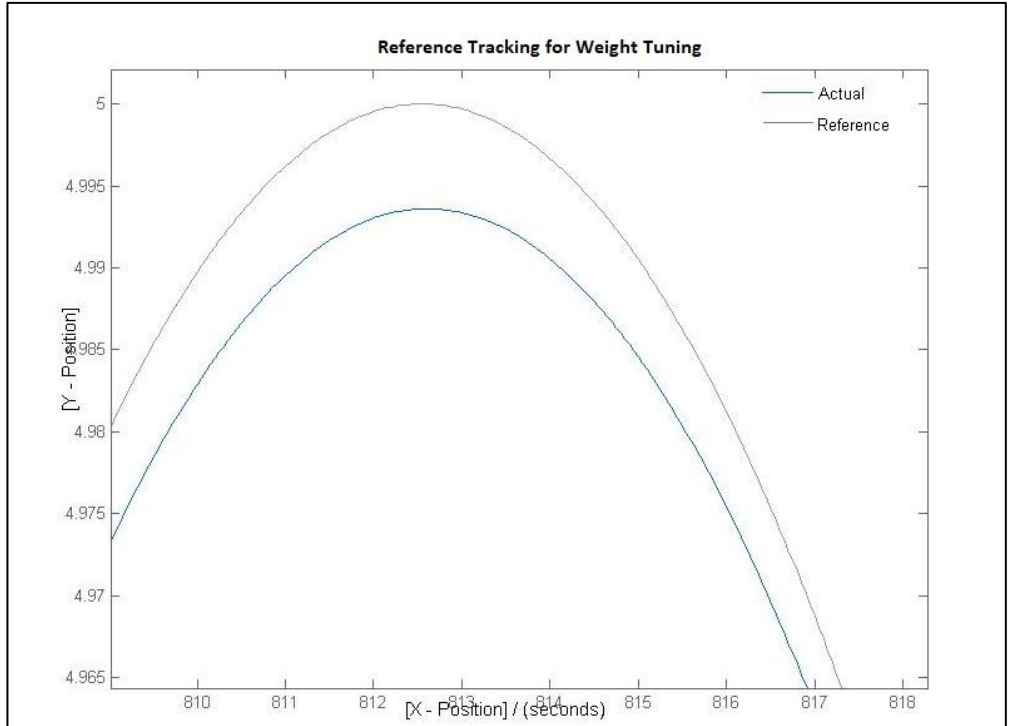

**Figure 8b: Weight Tuning MPC Response**

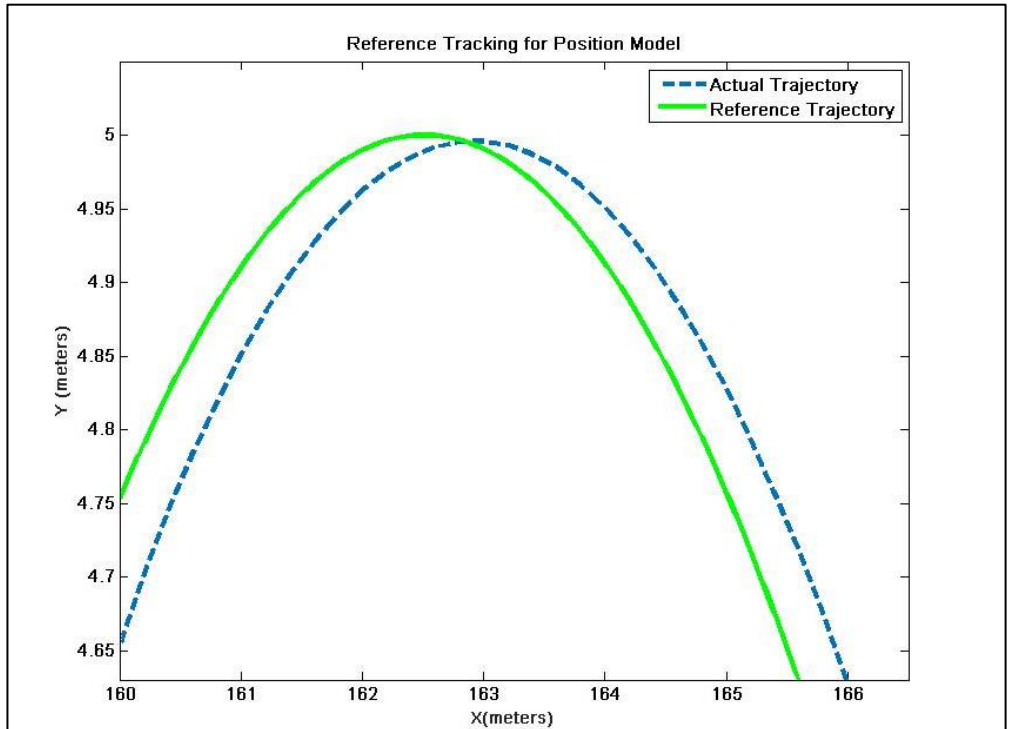

**Figure 8c: Reference Tracking for Position Model**

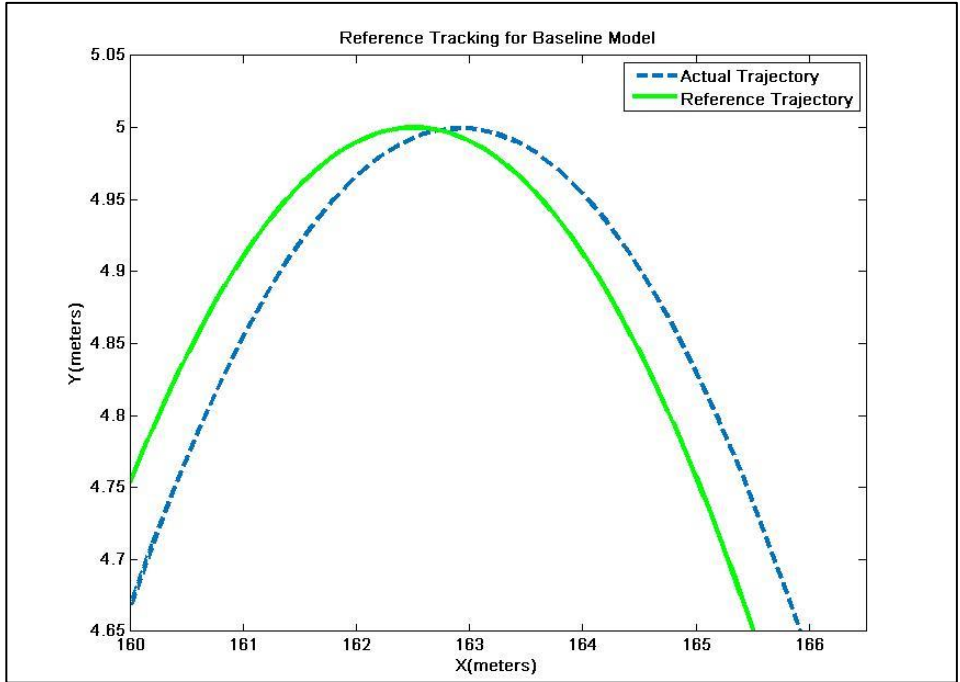
**Figure 8d: Reference Tracking for Baseline Model**

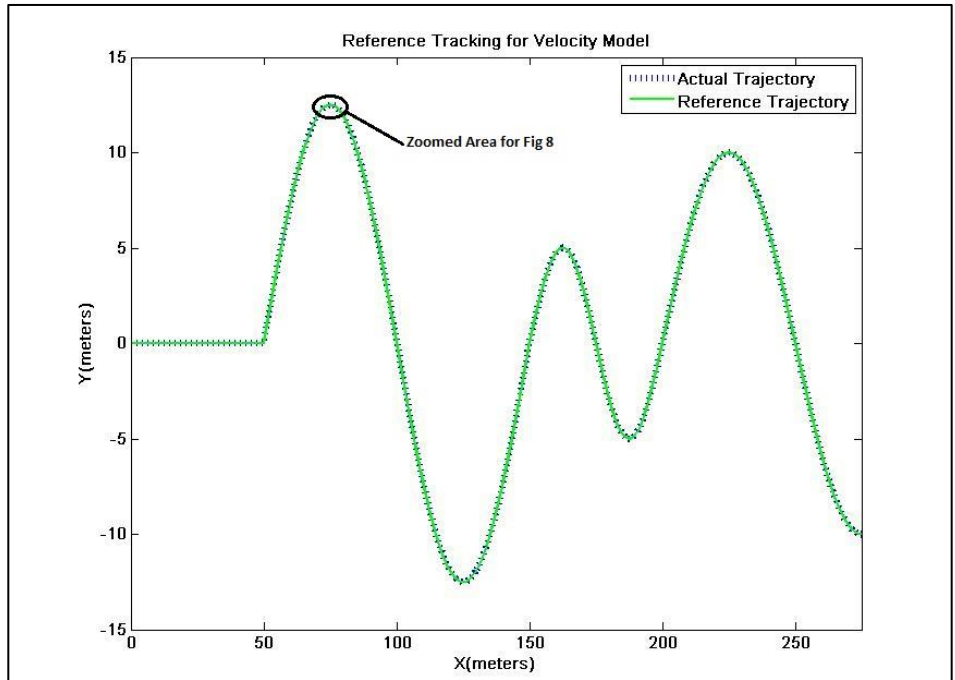
**Figure 9: Complete Trajectory**

# VIII. Conclusion

| Model | Baseline | Position | Velocity | Weight Tuning |
|---|---|---|---|---|
| Sum of Squared Difference ($m^2$) | 169.3724 | 2330 | 0.188 | 3.7215 |
| Time per Iteration (s) | 0.000545 | 0.001285 | 0.000232 | 8.33E-05 |
| Total Time (s) | 27.233 | 64.23862 | 11.6 | 4.165 |

Table 2: Table of SSD values and Computation Time

A table summarizing our results can be seen above in Table 2. We use the sum of squared difference (SSD) between the output and the reference trajectory to be tracked to evaluate the tracking performance of our models.

Comparing the results for constrained optimization, we find that the velocity model has the least SSD. The weight tuning method has the least computation time and the second smallest SSD. For our application of autonomous vehicles, depending on whether accuracy or speed is more important, one should pick the appropriate model, the velocity model has higher accuracy, but lower speed as compared to the weight tuning method and vice-versa. If safety is of greater importance, the weight tuning method would be favored.

The position model did not seem to perform as well as we anticipated, and we believe its performance can be better improved if more time is given.

## A. Future Work

Possible extensions of this work include trying to integrate the velocity model together with the weight tuning method and investigate its performance. Also, would be to compare the performance of nonlinear MPC with successive linearization method to see if there are computational cost-savings by solving a quadratic programming problem instead of a nonlinear optimization problem.

# IX. Contributions

| Name | Contributed Tasks |
|---|---|
| **Che Kun Law** | Adaptive Optimal Control [Position Model] |
| **Darshit Dalal** | Adaptive Optimal Control [Velocity Model] |
| **Stephen Shearrow** | Weight Tuning<br>Report Formatting |